\documentclass[letterpaper,12pt]{article}
\usepackage{tabularx} 
\usepackage{amsmath}
\usepackage{amsmath,bm}

\usepackage{graphicx} 
\usepackage[margin=1in,letterpaper]{geometry} 
\usepackage{cite} 
\usepackage[final]{hyperref} 
\hypersetup{
	colorlinks=true,       
	linkcolor=blue,        
	citecolor=blue,        
	filecolor=magenta,     
	urlcolor=blue         
}
\usepackage{amsfonts}
\usepackage{bbold}

\begin{document}
\title{Analytic approach to the plasmon hybridization in  rectangular nanoparticles}
\author{M. A. Kuntman,\; A. Kuntman and E. Kuntman}


\section*{Plasmon hybridization in rectangular nanoparticles}

M.A. Kuntman, A. Kuntman and E. Kuntman


\medskip
\medskip
\medskip
\medskip
\medskip
\noindent
\subsubsection*{Abstract}
Hybridized energies in rectangular nanoparticles may display an unusual behaviour. In some cases, they get separated with increasing distance between the particles. In this note this phenomenon is explained by an analytic method. 

\medskip
\medskip
\medskip

\subsubsection*{Scattering spectra of squares vs. rectangles}
\begin{figure}[h!]
\includegraphics[width=\linewidth,height=80pt]{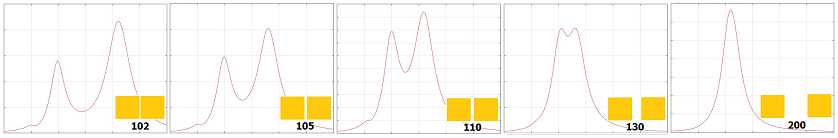}
\caption{\footnotesize{BEM simulations of scattering spectra for interacting squares. Particles are 100$\times$100 nm slabs with 10 nm thickness. Intensity is plotted versus wavelength for different distances ($d$) between the center of masses: $d=102$, 105, 110, 130, 200 nm   from left to right. Wavelength interval ranges from 500 to 1100 nm. } }
\label{squares}
\end{figure}

\begin{figure}[h!]
\includegraphics[width=\linewidth,height=80pt]{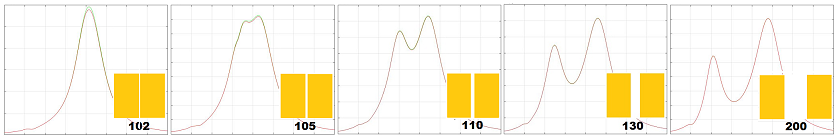}
\caption{\footnotesize{BEM simulations of scattering spectra for interacting rectangles. Particles are 180$\times$100 nm slabs with 10 nm thickness. Intensity is plotted versus wavelength for different distances ($d$) between the center of masses: $d=102$, 105, 110, 130, 200 nm   from left to right. Wavelength interval ranges from 500 to 1300 nm. }}
\label{rectangles}
\end{figure}

Plasmonic coupling between closely spaced nanoparticles results in the hybridized  resonance modes. In general, due to the weakening of dipole interaction,  hybridized modes (energies) come closer to each other with increasing distance between the particles.  Fig.\ref{squares} shows the scattering spectra of two interacting squares. However, in case of rectangular nanoparticles, for  a certain aspect ratio, hybridized energies may get
separated with increasing distance (Fig.\ref{rectangles}). 

\subsubsection*{Modeling} 
\begin{figure}[h!]
\centering
\includegraphics[width=140pt,height=56pt]{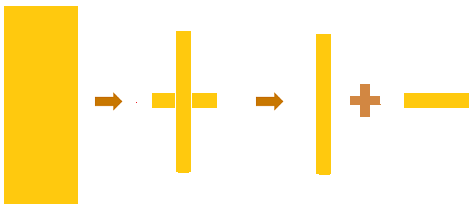}
\caption{\footnotesize{A rectangle can be modeled as two orthogonal rods.} }
\label{squaretocross}
\end{figure}

\begin{figure}[h!]
\centering
\includegraphics[width=238pt,height=56pt]{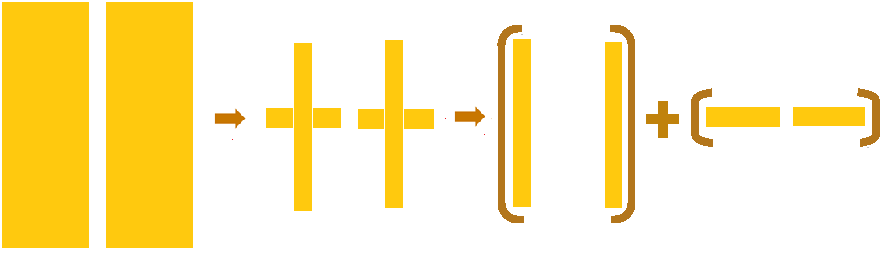}
\caption{\footnotesize{Coupled rectangles can be modeled as coupled crosses and two coupled crosses is equivalent to a system composed of two coupled vertical rods and two coupled horizontal rods.} }
\label{2squares}
\end{figure}

A rectangular slab can be modeled as two perpendicular nanorods that form a cross (Fig.\ref{squaretocross}). Two interacting rectangular slabs correspond to a pair of crossed nanorods (Fig.\ref{2squares}). 
A square slab is a special case with two identical nanorods
perpendicular to each other.

First, we consider a pair of crossed nanorods (Fig.\ref{2squares}). Because of the orthogonality there is no interaction between the vertical and horizontal components. The system can be viewed as a noninteracting combination of two coupled vertical nanorods and two coupled horizontal nanorods. It is worth noting that a rectangle is not equivalent to a cross. For instance, a rectangular slab of dimensions 180$\times$100 nm is not equivalent to a cross formed by 180 nm vertical and 100 nm horizontal rods. But, there exists a cross that spectrally mimics the  rectangle. As it can be seen in the following figures (Fig.\ref{equalcrosses} and Fig.\ref{unequalcrosses}) scattering spectra of coupled  180$\times$100 nm rectangles can be compared with that of coupled crosses formed by 135 nm vertical and 100 nm horizontal rods.   

Low energy mode of coupled vertical rods is dark and the bright mode drifts to lower energies (longer wavelengths) with increasing distance between the particles. But, as it will be discussed in the next section, interaction coefficient of vertical rods is small, and the energy shift is negligible for distances greater than 100 nm. On the other hand, high energy mode of coupled horizontal rods is dark and the bright mode moves to higher energies (shorter wavelengths) with increasing distance between the particles. Interaction coefficient of horizontal rods is more efficient, hence the energy shift is still prominent at distances greater than 100 nm.

\begin{figure}[h!]
\includegraphics[width=\linewidth,height=240pt]{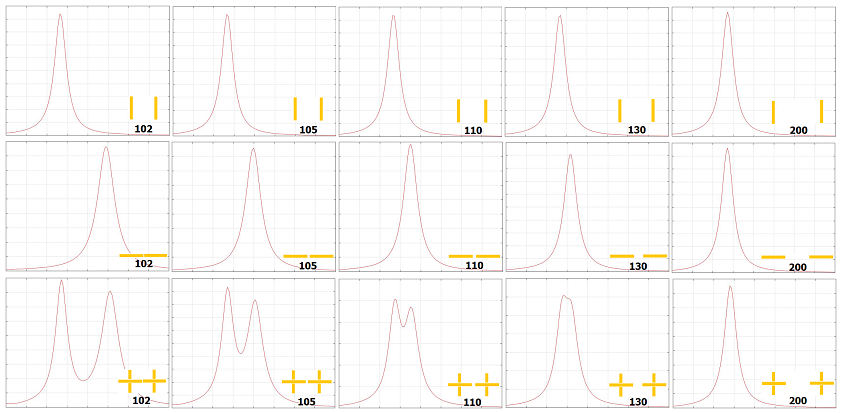}
\caption{\footnotesize{ First row: Two coupled 100$\times$10$\times$10 nm vertical rods. Second row: Two coupled 100$\times$10$\times$10 nm horizontal rods.  Third row: Two coupled crosses with equal arms. BEM simulations of scattering spectra are plotted versus wavelength in the interval 800-1600 nm. }}
\label{equalcrosses}
\end{figure}

\begin{figure}[h!]
\includegraphics[width=\linewidth,height=240pt]{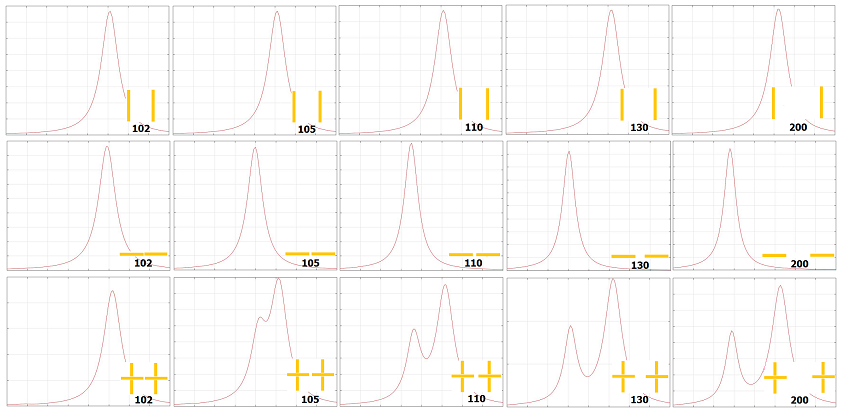}
\caption{\footnotesize{ First row: Two coupled 135$\times$10$\times$10 nm vertical rods. Second row: Two coupled 100$\times$10$\times$10 nm horizontal rods.  Third row: Two coupled crosses with unequal arms. BEM simulations of scattering spectra are plotted versus wavelength in the interval 800-1600 nm. }}
\label{unequalcrosses}
\end{figure}





Each row of Fig.\ref{equalcrosses} shows scattering spectra of two vertical rods (100 nm), two horizontal rods (100 nm) and a pair of crossed rods composed of them, respectively. At the beginning, bright mode of the vertical rods is at the left of the bright mode of the horizontal rods and they are well separated from each other (Fig.\ref{equalcrosses}). As the distance between the particles increases bright mode of vertical rods moves 
to the right (to the longer wavelengths) and the bright mode of the horizontal rods moves to the left. Combined effect can be seen in the last row of Fig.\ref{equalcrosses}. A pair of coupled crosses with equal arms have two bright modes due to the vertical and horizontal components, and they come closer with increasing distance between the crosses. Eventually, they merge at about $d=$200 nm.  


Each row of Fig.\ref{unequalcrosses} shows scattering spectra of two vertical rods (135 nm), two horizontal rods (100 nm) and a pair of crossed rods composed of them, respectively.  This configuration is designed on purpose to explain the unusual behaviour of rectangular particles. The length of the vertical rods is chosen longer than the length of the horizontal rods such that bright modes of vertical and horizontal components  overlap at the beginning ($d=$102 nm). As the distance between the particles increases bright mode of the coupled vertical rods moves to the right (to the longer wavelengths) and the bright mode of the coupled horizontal rods moves to the left (to the shorter wavelengths) as before. Therefore, it is not a surprise to observe that, in case of coupled pair of crosses with unequal arms (third row of Fig.\ref{unequalcrosses}), hybridized modes  separate more from each other with increasing distance.


\subsubsection*{Jones matrix of rectangles}
For simple nano systems like spherical particles, it is enough to investigate the behavior of the system under a single polarization excitation. However, in case of coupled rectangular particles, one usually needs to study different excitation polarizations, and matrix methods that employ all possible excitation modes become important\cite{Thesis,OA}.

Nanorods are the basic elements of a class of more complex nanosystems. Their optical response can be modeled as oriented dipoles with polarization characteristics similar to that of linear polarizers in a certain interval of photon energy. We assume that the polarizability of the rod is fully anisotropic, i.e., the particle can only polarize along a particular direction.
Hence, scattering properties of a nanorod oriented in a certain direction in  space can be represented by a linear polarizer Jones matrix:

\begin{equation}
\mathbf{J}= \alpha  \left(
\begin{array}{cc}
\cos^2\theta & \cos\theta\sin\theta \\
\cos\theta\sin\theta & \sin^2\theta \\
\end{array}
\right)
\end{equation}
where $\alpha$
is the Lorentzian polarizability associated with the nanorod, $\theta$ is the orientation angle in the $x-y$ plane measured from the $x$ axis.

Jones matrix of a cross formed by two equal nanorods is a $2\times 2$ identity matrix with a Lorentzian polarizability that has a resonance at frequency $\omega_0$ \cite{VMS}-\cite{Young}:

\begin{equation}
\mathbf{J}=\alpha\begin{pmatrix}
1&0\\0&1
\end{pmatrix}
\end{equation}
Jones matrix of a square slab is also an identity matrix with a Lorentzian polarizability of its own. In case of a cross made of unequal rods the Jones matrix ($\mathbf{J}_C$) has two different polarizabilities:

\begin{equation}\label{Jcross}
\mathbf{J}_C=\begin{pmatrix}
\alpha_H&0\\0&\alpha_V
\end{pmatrix}
\end{equation}
where $\alpha_H$ and  $\alpha_V$ are the polarizabilities for the vertical and horizontal rods with resonance frequencies $\omega_H$ and $\omega_V$, respectively.
Similarly, according to the modeling presented in the previous section, Jones matrix of a rectangle ($\mathbf{J}_R$) has two polarizabilities $\alpha'_H$ and $\alpha'_V$ with resonance frequencies $\omega'_H$ and $\omega'_V$:

\begin{equation}\label{Jrectangle}
\mathbf{J}_R=\begin{pmatrix}
\alpha'_H&0\\0&\alpha'_V
\end{pmatrix}
\end{equation}

\begin{figure}[h!]
\centering
\includegraphics[width=300pt,height=80pt]{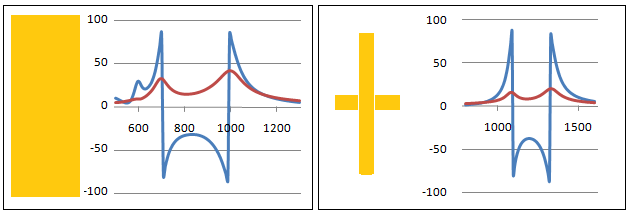}
\caption{\footnotesize{Left panel; angle $\phi'$ is plotted vs wavelength for a 180$\times$100$\times$10 nm rectangle. Right panel; angle $\phi$ for a cross made of 135$\times$10$\times$10 nm vertical rod and 100$\times$10$\times$10 nm horizontal rod. Intensities (in red) are shown for reference}}
\label{aciveintens}
\end{figure}

Jones matrix of a cross with unequal arms (or a rectangle) is as a retarder matrix, in general. It is useful to write Eqs.\eqref{Jcross} and \eqref{Jrectangle} as follows:
\begin{equation}
\mathbf{J}_C=\alpha_H\begin{pmatrix}
1&0\\0&\rho e^{i\phi}
\end{pmatrix}, \quad \mathbf{J}_R=\alpha'_H\begin{pmatrix}
1&0\\0&\rho' e^{i\phi'}
\end{pmatrix}
\end{equation}
where $\rho e^{i\phi}=\alpha_V/\alpha_H$ ($\rho' e^{i\phi'}=\alpha'_V/\alpha'_H$). Fig. \ref{aciveintens} shows variation of $\phi$ and $\phi'$ with respect to the wavelength. Graphics are obtained from the BEM simulations. Intensity peaks (hybridized energies) coincide with the inflection points of $\phi$ ($\phi'$). These figures show that a rectangle and a cross with unequal arms have very similar characteristics.  
\subsubsection*{Analytic approach to interacting rectangles}
The induced electric dipole moment vector, $\mathbf{P}$, on a nanorod is proportional to the incident electric field, $\mathbf{E}_0(\mathbf{r})$\cite{PRB}:
\begin{equation}
\mathbf{P}=\varepsilon\mathbf{J}\mathbf{E}_0(\mathbf{r}),
\end{equation}
where $\varepsilon$ is  the permittivity of the medium at the dipole position. 

When we put two particles close to each other we have to consider mutual interaction contributions. Each one of the induced dipoles will experience the field of the
other dipole. This coupling effect can be taken into account to find the actual dipole of each particle as follows: 
\begin{subequations}
\begin{equation}\label{dipole1}
\mathbf{P}_{1}= \mathbf{J}_{1}[\varepsilon 
   \mathbf{E_o}(\mathbf{r}_{1}) + k^2 \mathbf{\bar{\bar{G}}}(\mathbf{r}_{1}-\mathbf{r}_{2})\cdot \mathbf{P}_{2}],
\end{equation}
\begin{equation}\label{dipole2}
\mathbf{P}_{2}= \mathbf{J}_{2}[\varepsilon 
   \mathbf{E_o}(\mathbf{r}_{2}) + k^2 \mathbf{\bar{\bar{G}}}(\mathbf{r}_{2}-\mathbf{r}_{1})\cdot \mathbf{P}_{1}],
\end{equation}
\end{subequations}
where $k$ is the wavenumber, $\mathbf{J}_{1}$, $\mathbf{J}_{2}$ are the Jones matrices of the particles and $\mathbf{\bar{\bar{G}}}$ is the free-space electric dyadic Green's function with the following effect on the dipole vector \cite{14}:

\begin{equation}
\begin{split}
\mathbf{\bar{\bar{G}}}\cdot \mathbf{P} =& \bigg[\left(1 + \frac{i}{kd} - \frac{1}{k^2d^2}\right) \mathbf{P}  \\ &+ \left(-1 - \frac{3i}{kd} + \frac{3}{k^2d^2}\right)(\mathbf{\hat{u}}\cdot\mathbf{P})\mathbf{\hat{u}}\bigg]g(d),
\end{split}
\end{equation}
where $g(d)=e^{ikd}/4\pi d$, $d$ is the distance and $\mathbf{\hat{u}}$ is the unit vector between the center of masses of particles. The notation can be simplified if we let,
\begin{subequations}
\begin{equation}
A=\left(1 + \frac{i}{kd} - \frac{1}{k^2d^2}\right)g(d),
\end{equation}
\begin{equation}
B= \left(-1 - \frac{3i}{kd} + \frac{3}{k^2d^2}\right)g(d),
\end{equation}
\end{subequations}
thus,
\begin{equation}
\mathbf{\bar{\bar{G}}}\cdot \mathbf{P} = A \mathbf{P} + B(\mathbf{\hat{u}}\cdot\mathbf{P})\mathbf{\hat{u}}.
\end{equation}

There are two equivalent approaches to the issue of plasmon hybridization in coupled pair of crosses. It is possible to consider a pair of crosses as a system consisting of four interacting nanorods: two vertical and two horizontal nanorods with the associated vertical and horizontal linear polarizer Jones matrices; or, consider each cross as a single subsystem with the Jones matrix given by Eq.\eqref{Jcross}.

We prefer the second approach and apply Eqs.\eqref{dipole1} and \eqref{dipole2}. After solving them we get two coupled equations for the horizontal components of the dipole vectors ($P_{1x}$ and $P_{2x}$) and two coupled equations for the vertical components of the dipole vectors ($P_{1y}$ and $P_{1y}$). Equations for horizontal components are not coupled to the equations for vertical components because of the orthogonality, hence, they can be solved easily:
\begin{subequations}
\begin{equation}\label{x}
P_{1x}=P_{2x}=\frac{\varepsilon\alpha_H  E_{0x}(1+\alpha_H\delta_2)}{1-\alpha^2_H\delta^2_2}
\end{equation}
\begin{equation}\label{y}
P_{1y}=P_{2y}=\frac{\varepsilon\alpha_V  E_{0y}(1+\alpha_V\delta_1)}{1-\alpha^2_V\delta^2_1}
\end{equation}
\end{subequations}
where $E_{ox}$, $E_{oy}$ are the components of the incident electric field, and $\delta_1, \delta_2$ are the interaction coefficients defined as follows \cite{PRB}:
\begin{equation}
\delta_1=k^2A, \quad \delta_2=k^2(A+B)
\end{equation}
Peak values of the spectra should be determined from the extremum points of the denominators of Eqs.\eqref{x} and \eqref{y}. Quadratic form of the denominators suggest two hybridized energies \cite{PRB}. But, one of the modes is dark, because Eqs.\eqref{x} and \eqref{y} can be simplified as follows:
\begin{subequations}
\begin{equation}\label{x2}
P_{1x}=P_{2x}=\frac{\varepsilon\alpha_H  E_{0x}}{1-\alpha_H\delta_2}
\end{equation}
\begin{equation}\label{y2}
P_{1y}=P_{2y}=\frac{\varepsilon\alpha_V  E_{0y}}{1-\alpha_V\delta_1}
\end{equation}
\end{subequations}
Therefore, $x$ components of the electric dipole vectors has a bright mode at the extremum point of the denominator of Eq.\eqref{x2} which is determined by $\alpha_H$ and $\delta_2$; $y$ components of the electric dipole vectors has a bright mode at the extremum point of the denominator of Eq.\eqref{y2} which is determined by $\alpha_V$ and $\delta_1$.

\subsubsection*{Jones matrix for the coupled rectangles}


Jones matrix of the coupled crosses (rectangles) can be directly written from the far field effects of the dipole fields:

\begin{equation}\label{Jcoupled}
\mathbf{J}=2\varepsilon\beta\begin{pmatrix}\frac{\alpha_H}{1-\alpha_H\delta_2}&0\\0&\frac{\alpha_V}{1-\alpha_V\delta_1}\end{pmatrix}
\end{equation}
where $\beta$ is the far field coefficient \cite{PRB}.
Eq.\eqref{Jcoupled} is a linear combination of two Jones matrices corresponding to horizontal and vertical linear polarizer matrices ($\mathbf{J}_H$ and $\mathbf{J}_V$): 

\begin{equation}\label{Jsum}\mathbf{J}=\mathbf{J}_H+ \mathbf{J}_V=\frac{2\varepsilon\beta\alpha_H}{1-\alpha_H\delta_2}\begin{pmatrix}1&0\\0&0\end{pmatrix}+\frac{2\varepsilon\beta\alpha_V}{1-\alpha_V\delta_1}\begin{pmatrix}0&0\\0&1\end{pmatrix}
\end{equation}
It is worth noting that, although $\mathbf{J}$ is written as a combination of two linear polarizer matrices, it is not a linear polarizer Jones matrix, in general \cite{PRA}.

Denominators result from the hybridization. If, we retain only the near-field contribution, i.e., retain only the
$1/d^{3}$ term in $A$ and $B$, the interaction coefficients  simplify to
 $\delta_1 = -1/4\pi d^3$ and $\delta_2 = 2/4\pi d^3$ \cite{43}.  
Magnitude of $\delta_2$ is 2 times greater than $\delta_1$. This explains why the coupling between the horizontal components is more prominent. According to the near field approximation the Jones matrices  of the coupled vertical and horizontal rods can be written,
respectively, as follows:

\begin{equation}
\mathbf{J}_H=\frac{2\varepsilon\beta\alpha_H}{1-2\alpha_H/4\pi d^3}\begin{pmatrix}1&0\\0&0\end{pmatrix}, \quad  \mathbf{J}_V=\frac{2\varepsilon\beta\alpha_V}{1+\alpha_V/4\pi d^3}\begin{pmatrix}0&0\\0&1\end{pmatrix}
\end{equation}
For $\mathbf{J}_H$, the bright (in-phase) mode occurs at the energy that makes  $Re(2\alpha_H/4\pi d^3)=1$, and moves to the higher energies with increasing distance. Meanwhile, for $\mathbf{J}_V$, the bright (in-phase) mode occurs at the energy that makes $Re(\alpha_V/4\pi d^3)=-1$, and moves to the lower energies with increasing distance. In both cases, out-of-phase mode is dark, because it cannot be optically activated with a plane wave.



\end{document}